\documentclass[%
 aapm,
 mph,%
 amsmath,amssymb,
preprint,%
]{revtex4-2}

\usepackage{graphicx}
\usepackage{dcolumn}
\usepackage{bm}

\begin{document}

\title{Spin-Orbit Interactions of Light: Fundamentals and Emergent Applications}%

\author{Graciana Puentes$^{1,2}$}%
\affiliation{1-Departamento de Fsica, Facultad de Ciencias Exactas y Naturales,
Universidad de Buenos Aires, Ciudad Universitaria, 1428 Buenos Aires, Argentina\\
2-CONICET-Universidad de Buenos Aires, Instituto de Fsica de Buenos Aires (IFIBA), Ciudad Universitaria,
Buenos Aires, Argentina.
}%

\date{\today}

\begin{abstract}
We present a comprehensive review of recent developments in Spin Orbit Interactions (SOIs) of light in photonic materials. In particular, we highlight progress on detection of Spin Hall Effect (SHE) of light in hyperbolic metamaterials and metasurfaces. Moreover, we outline some fascinating future directions for emergent applications of SOIs of light in photonic devices of the upcoming generation.
\end{abstract}

\maketitle


\section{Introduction}

Light's polarization degrees of freedom, also known as Spin Angular Momentum (SAM) and orbital degrees of freedom, also known as Orbital Angular Momentum (OAM), can be coupled to produce a wide variety of phenomena, known as Spin-Orbit Interactions (SOIs) of light. Due to their fundamental origin and diverse character, SOIs of light have become crucial to a variety of active fields, such as singular optics, photonics, nano-optics and quantum optics, when dealing with SOIs at the single-photon level. Among a variety of fascinating exotic phenomena, SOIs show the remarkable spin-dependent transverse shift in light intensity, also known as the Spin Hall Effect (SHE) and the Spin Orbit Conversion (SOC) of light. \\

The various plane-wave components of the beam that travel in slightly different directions and acquire slightly different complex reflection or transmission coefficients are what cause the regular SHE of light at a planar interface to form [1-4]. A handy quantum-like framework with generalized wavevector-dependent Jones-matrix operators at he interface, and expectation values of the position and momentum of light, provides a theoretical explanation of the photonic  SHE [3-6]. Such a description also unifies the longitudinal (in-plane) beam shifts connected to the Goos—Hanchen (GH) effect [3-6] and transverse SHE shifts, also known as the Imbert—Fedorov (IF) shifts in the case of the Fresnel reflection/refraction [7-10].

2D metamaterials, commonly referred to as metasurfaces, are a an interdisciplinary area that encourages the use of substitute
methods for light engineering based on spatially ordered meta-atoms and subwavelength-thick metasurfaces of different compositions. They display exceptional qualities in
light manipulation in a 2D interphase. Metasurfaces can attain their 3D counterpart functions, such as invisibility cloaking and negative refractive index. In addition, they can eliminate some of the 3D metamaterial current restrictions, 
such as high resistivity or dielectric loss, for example. Additionally, the creation of metasurfaces via
conventional methods for nanofabrication, such as electron beam lithography techniques, are readily available in the semiconductor industry.\\

In this review, we provide a summary of recent findings and future potentials for applications of SHE of light in photonic
materials. As an optical equivalent of the solid-state spin Hall effect,  SHE of light warrants promissing opportunities for examination of innovative photonic materials and nanostructures physical characteristics, such as in figuring out the magnetic and metallic thin films' material characteristics, or the optical characteristics of two-dimensional atomically thin
metamaterials, with unmatched spatial and angular precision resolution, a trait that SHE and other combined technologies can provide, utilizing quantum weak measurements and quantum
weak amplification methods. Additionally, we provide a summary of recent developments in
 primary 2D metamaterials and metasurfaces
applications for producing and manoeuvering Spin Angular Momentum (SAM) and
Orbital Angular Momentum (OAM) of light, for applications in multicasting and multiplexing, spin-based metrology or quantum networks.\\

\section{Spin Hall Effect of Light (SHEL)}

Using the terminology of the closely related work, we begin with the theoretical description of the problem. Namely, the SHE of light in a tilted photonic material. Figure 1 (a), (b) and (c) depict the problem's geometry. The normalized Jones vector describes the incident $z$-propagating paraxial beam's polarization $\left| \psi  \right\rangle = (E_x ,E_y)^{T}$ ($T$ stands for the transposition operator), $\left\langle \psi \! \right.\left|\, \psi \right\rangle = |E_x|^2 + |E_y|^2 =1$. In the $(x,z)$ plane, the photonic material is tilted so that its axis forms a $\theta$ angle (labeled $\vartheta$ for ease of notation in the following sections) with the $z$-axis and transmits mostly the $y$-polarization. The dichroic action of the photonic material in this geometry, and in the zero-order approximation of the incident plane-wave field, can be characterized by the Jones matrix, of the form:

%
\begin{equation}
\hat{M}_{0}=\left( \begin{array}{cc}
T_{x}(\theta) & 0  \\
0 & T_{y}(\theta) \end{array} \right),
\end{equation}
%

the Jones vector of the transmitted wave is $\left| \psi' \right\rangle = \hat{M}_0 \left| \psi \right\rangle$. $T_{x,y}$, which can depend on $\theta$, are the amplitude transmission coefficients for the $x$- and $y$-polarized waves. While $T_{x }= 0$ and $T_{y} = 1$ for an ideal polarizer, we can assume that $|T_x/T_y| \ll 1$ for real dichroic plates. Also take note of the fact that $T_{x,y} = \exp (\mp i\Phi/2)$  relates to the birefringent waveplate issue discussed in [11].

Considering that in the paraxial approximation the beam consists of a superposition of plane waves with their wavevector directions labelled by small angles ${\bm \Theta} = \left(\Theta_x,\Theta_y \right)\simeq \left(k_x/k,k_y/k \right)$ [see Fig.~1(a)], the Jones matrix incorporates $\bm \Theta$-dependent corrections and can be wrtten as [11]:
%

\begin{equation}
\hat{M}(\bm \Theta)=\left( \begin{array}{cc}
T_{x}(1 + \Theta_{x}X_{x}) & T_{x}\Theta_{y}Y_{x} \\
-T_{y}\Theta_{y}Y_{y} &T_{y}(1 + \Theta_{x}X_{y})  \end{array} \right),
\end{equation}

%
here
\begin{equation}
X_{x,y}= \frac{d \mathrm{ln}T_{x,y}}{d \theta }  \hspace{1cm}  Y_{x,y}= (1- \frac{T_{y,x}}{T_{x,y}} )  \mathrm{cot}(\theta),
\end{equation}

%
%
are the well-known Goos-Hanschen (GH) and Spin Hall (SHEL) terms [11], the latter being the main focus of this review. 

\section{Quantum Weak Amplification Techniques}

\begin{figure}[b!]
\centering
\includegraphics[width=0.9\linewidth]{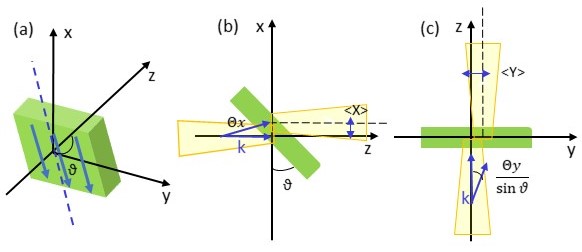}
\caption{(a) General 3D geometry of the problem displaying the angle $\vartheta$ between the anisotropy axis of the plate and the beam axis $z$. (b) The wave vectors in-plane  deflections $ (\Theta_{x})$ cause the well-known birefringence shift $\langle \hat{X} \rangle$ which is comparable to the GH shift, by altering the angle between k and the anisotropy axis. (c) View along the anisotropy axis of the crystal is shown. The transverse $\Theta_{y}$ deflections of the wave vectors rotate the corresponding planes of the wave propagation with respect to the anisotropy axis by the angle $\vartheta \approx \Theta_{y}/\mathrm{sin} (\Theta_{y}) $. This causes a new helicity-dependent transverse shift $\langle \hat{Y} \rangle$, i.e., a spin-Hall effect similar to the IF shift. Further details are in the text.}
\label{fig:false-color}
\end{figure}

A detailed explanation of the entire theory governing transverse shifts of light in tilted uniaxial crystals was ellaborated in [11,12,13]. The anisotropic transverse shift, or SHE of light, is represented by the expectation value of the position operator $\langle \hat{Y} \rangle$, for an input state defined by a Jones vector $|\psi \rangle$, and a transmitted state described by a Jones vector $|\psi' \rangle$:

\begin{equation}
\langle \hat{Y} \rangle= \langle \psi' | \hat{Y}| \psi \rangle=\frac{cot(\vartheta)}{k}[-\sigma (1-\cos( \phi_0)) + \chi \sin(\phi_0) ],
\end{equation}

\noindent where $k$ is the wave-vector, $\phi_0$ refers to the difference in phase between ordinary and extraordinary waves as they propagate through a birefringent material, $\vartheta$ represents the tilt angle of the photonic medium, and $(\sigma, \chi, \tau)$ represent the Stokes parameters for the light beam. The SHE can be measured directly, using the sub-wavelength shift of the beam centroid [14-18] [Eq. (3)]. Various alternative techniques, such as quantum weak measurements [19-27] are also used. The latter technique enables substantial amplification, when employing almost crossed polarizers for pre-selection and post-selection of the polarization input and output states, as they relate to the system. \\

The output polarizer corresponds to a post-selected polarization state $|\psi \rangle=(\alpha',\beta')^{T}$  whereas the input polarizer corresponds to a pre-selected state  $|\psi \rangle=(\alpha,\beta)^{T}$, here $(\alpha(\alpha ' ),\beta(\beta '))$ represent the input(output) polarization components of the input(output) beam ($E_x(E_x'),E_y(E_y'))$, and $T$ stands for transposition operation. As opposed to traditional expectation values, weak values $ \langle \hat{Y} \rangle_{weak}$ might display a quantum weak amplification effect and lie beyond the operator's spectrum, moreover weak values can take imaginary values. We examine the quantum weak amplification of the SHE shift using an initial beam with $e$ polarization $|\psi\rangle=(1,0)^{T}$, and a nearly orthogonal polarization state $|\phi \rangle=(\epsilon,1)^{T}, |\epsilon|<<1$, for the post-selection polarizer. In this configuration, the SHE weak value results in:

\begin{equation}
\langle \hat{Y} \rangle_{weak}=\frac{1}{\epsilon k}\sin(\phi_0) \cot(\vartheta)+\frac{z}{z_{R}}\frac{1}{\epsilon k}(1-\cot(\phi_0))\cot(\vartheta),
\end{equation}

\noindent here $z_{R}$ stands for the Rayleigh length. The second angular term, becomes dominant in the far field zone, and presents weak amplification due to two reasons: First, due to the fact that $|\epsilon|<<1$, and second due to the fact that $z >> z_{R}$, in the far field regime. It should be noted that the maximum weak amplification that can be achieved at $|\epsilon| \approx (k \omega_0)^{-1}$ is of the order of the beam waist $\omega_0 z/z_{R}$.\\

 We compute the expected value of the centroid displacement based on the phase difference aquired by the beam (Eq. 3) as it propagates through the photonic material to estimate the classical beam shift. This is how the phase difference accumluated during propagation in the photonic material is expressed:\\

\begin{equation}
\phi_0(\vartheta)=k[n_ {o} d_{o}(\vartheta)-\bar{n}_{e}(\vartheta)d_{e}(\vartheta)],
\end{equation}

where,  $n_{e}(\vartheta)=n_{o}n_{e}/\sqrt{n_{e}\cos(\vartheta)+n_{o}\sin(\vartheta)}$ is the refractive index for
the extraordinary wave, and $n_{o}$  is the refractive index for the ordinary wave, which propagate at the angle $\vartheta$ to the optical
axis. The distances of propagation of the ordinary and extraordinary rays in the tilted plate are:

\begin{equation}
d_{e}(\vartheta)=\frac{\bar{n}_{e}(\vartheta)d}{\sqrt{\bar{n}_{e}^2(\vartheta) - \cos(\vartheta)^2}}, \hspace{0.3cm} d_{o}(\vartheta)=\frac{n_o d}{\sqrt{n_{o}^2-\cos(\vartheta)^2}}.
\end{equation}

Using Eqs. (3), (5) and (6) the expectation value of the spin Hall shift $\langle \hat{Y} \rangle$ and its weak value $\langle \hat{Y} \rangle_{weak}$ can be derived, and contrastrasted with experimental findings. 

\section{SAM, OAM and TAM of Light }
\label{sec3}
\vspace{-6pt}

\subsection{SAM of Light}

In optics, the Spin Angular Mometum (SAM) and Orbital Angular Momentum (OAM) of light can be observed separately. A Spin-Orbital Angular Momentum decomposition for paraxial monochromatic beams is simple. This distinctive property, which inspires this review, explains in part the recent unrivaled development in photonic SAM-OAM conversion in metasurfaces and 2D metamaterials. At the same time, the spin and orbital description of quantum electromagnetic field theories produce a variety of complexities, in the generic non-paraxial or non-monochromatic angular momentum description [10,11]. 

The SAM is connected to the polarization of light in the unified theory of angular momentum of light, which is based on canonical momentum and spin densities developed in [11]. Accordingly, right-hand circular (RHC) and left-hand circular (LHC) polarizations of a paraxial beam correspond to positive and negative helicity $ \sigma = \pm 1$. If the beam's mean momentum (measured in $\hbar$ units per photon) can be connected to its mean wave vector $\langle \textbf{k} \rangle$, then such beam carries the corresponding SAM $\langle S \rangle = \sigma \langle \bf{k} \rangle/| k| $, 
where the helicity parameter $\sigma$ is equivalent to the degree of circular polarization in the Jones formalism. 

A plane wave is an idealized phenomenon that can extend to the infinity. Extrinsic OAM cannot be carried by such a plane wave (like its mechanical counterpart $\bf{L}={r} \times {p}$),
because its position $\bf{r}$ is undefined. On the other hand, a circularly-polarized electromagnetic plane wave can carry SAM. In the canonical momentum representation, the vector describing the electric field, for a circularly polarized plane-wave propagating along the \emph{z}-direction can be written as \cite{Bliokh2015}:

\begin{equation}
\bf{E} \propto \frac{\hat{x} + i \sigma \hat{y}}{2} \exp(i|k|z), 
\end{equation}
where ($\hat{x} ,\hat{y} ,\hat{z}$) are unit vectors and the helicity parameter $\sigma  = \pm 1$ corresponds to the LHC and RHC polarizations, respectively. The wave number $|k|$, results from the dispersion relation for a plane wave, that is, $ |k| = \omega / c$. The electric field described in Equation (8) represents the eigen-mode of the \emph{z}-component of the spin-1 matrix operators with eigenvalue $\sigma$, of the form $\hat{S}_{z} \bf{E} = $ $\sigma$ ${\bf{E}}$  \cite{Bliokh2015}. Where the spin-1 operators (generators of the SO(3) vector rotations)  are given by \cite{Bliokh2015}:

\begin{equation}
\hat{S}_{x}=-i\left( \begin{array}{ccc}
0 & 0 & 0 \\
0 & 0 & 1 \\ 
0 & -1 & 0 \end{array} \right), \hspace{3mm} \hat{S}_{y}=-i\left( \begin{array}{ccc}
0 & 0 & -1 \\
0 & 0 & 0 \\
1 & 0 & 0 \end{array} \right), \hspace{3mm} \hat{S}_{z}=-i\left( \begin{array}{ccc}
0 & 1 & 0 \\
-1 & 0 & 0 \\
0 & 0 & 0 \end{array} \right). 
\end{equation}

Therefore, the plane wave carries SAM density $\langle S  \rangle = \sigma \frac{\langle \bf{k} \rangle}{|\bf{k}|}$, defined as the local expectation value of the spin operator with the electric field Equation (8).

\subsection{OAM of Light}

In 1936, Beth made the first demonstration of the mechanical torque produced by the transmission of angular momentum from a circularly polarized light beam to a birefringent plate [30,31]. In this experiment, a  fiber suspended a fine quartz quarter-wave plate. Such plate transforms RHC polarization, with spin component $+\hbar$, into LHC polarization, with spin component $-\hbar$, with a net SAM transfer of $2 \hbar$ per photon to the birefringent plate. Beth measured torque, known as the measurement of SAM of the photon, agreed in sign and modulus with the quantum and classical expectations.

In Reference [31], Laguerre-Gaussian modes with azimuthal angular dependency ($\exp{[-il\phi]}$) were shown to exist, which are {eigen}-modes of the momentum operator $L_{z}$ and carry an orbital angular {momentum} $l \hbar$ per photon. Using the vector potential, a proper representation of a linearly polarized TEM$_{plq}$ laser mode can be obtained, in the Lorentz gauge [30]:

\begin{equation}
A=u(x,y,z)\exp{[-ikz]}\hat{x},
\end{equation}
where $u(r,\phi,z)$ is the complex {scalar} field amplitude satisfying the paraxial wave equation and $\hat{x}$ is the unit vector in \emph{x}-direction. In the paraxial regime,  $du / dz$ is taken to be small compared to $ku$ and second derivatives and the products of first derivatives of the electro-magnetic field are ignored. The solutions describing the Laguerre-Gauss beam, for the cylindrically symmetric case $u_{r,\phi,z}$,  are of the form [31]:
\begin{eqnarray}
u_{r,\phi,z}&=& \frac{C}{\sqrt{1+\frac{z^2}{z_{R}^2}}}[\frac{r \sqrt{2}}{w(z)}]^l L_{p}^{l} [\frac{2r^2}{w(z)^2}]\\ \nonumber
                  &   &  \times\exp{[\frac{-r^2}{w(z)^2}]}\exp{[\frac{_ikr^2z}{2(z^2+z_{R}^2}]} \exp{[-il\phi]}\\ \nonumber
                    &  &  \times \exp{[i(2p+l+1)\tan^{-1}[\frac{z}{z_{R}}]]},
\end{eqnarray}
where $w(z)$ is the radius of the beam, $L_{p} ^{l}$ is the associated Laguerre polynomial, $z_{R}$ is the Rayleigh range, $C$ is a constant and the beam waist is taken at $z=0$. Within this description, the time avarage of the real part of the {Poynting} vector $\epsilon_0 E \times B$, results in:

\begin{equation}
\frac{\epsilon_0}{2} (E ^{*} \times B + E \times B^{*}) = i \omega  \frac{\epsilon_0}{2} (u^{* }\nabla u - u \nabla u ^{*}) + \omega k \epsilon_0 |u|^2 \hat{z},
\end{equation}
where $z$ is the unit vector in \emph{z}-direction. When applying to a Laguerre-Gaussian distribution given in Equation (4), the linear momentum density takes the form:

\begin{equation}
P=\frac{1}{c} [\frac{rz}{z^2+ z_{R} ^2}|u|^2 \hat{r} + \frac{1}{kr}|u|^2 \hat{\phi} + |u|^2 \hat{z}],
\end{equation}
where $\hat{r}, \hat{\phi}$ are unit vectors. It may be seen that the Poynting vector ($c^2 P$) spirals along the direction of propagation along the beam. The $z$ component relates to the linear momentum, the $r$ component to the spatial dispersion of the beam, and the $ \phi$ component generates the OAM.

\section{Photonic Materials: Metamaterials and Metasurfaces}

\begin{figure}[b!]
\centering
\includegraphics[width=0.8\textwidth]{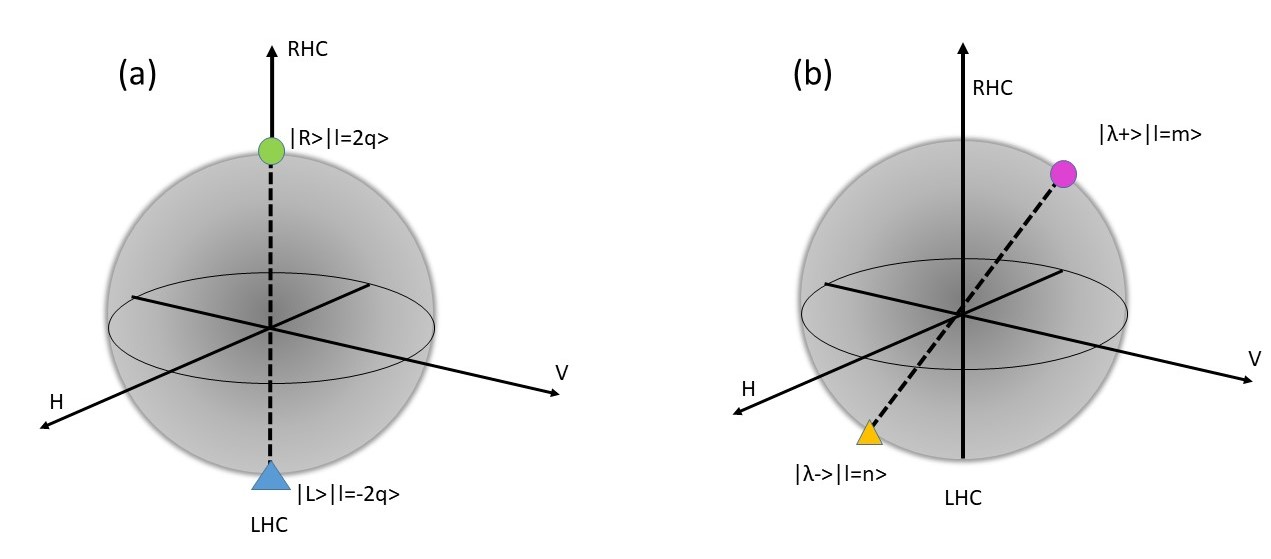}
\caption{(\textbf{a}) Representation in Poincare sphere of Spin-Orbit-Conversion (SOC) via $q$-plates. A state of circular polarization located in the poles of the Poincare sphere is mapped into its opposite state of circular polarization, while imprinting a fixed {azimuthal} phase $\pm 2q \phi$. {(\textbf{b})} Representation in Poincare sphere of spin-orbit {conversion} (SOC) via $J$-plates. Arbitrary states of elliptic polarization $ | \lambda \pm \rangle$ are mapped into its opposite state of  elliptic polarization, while imprinting a tunable {azimuthal} phase~$ (n,m)\phi$. }
\end{figure}

\subsection{SAM Control in Metasurfaces}

Metasurfaces enable polarization conversion in the same way as ordinary wave plates by maneuvering two light eigenmodes that correspond to orthogonal polarization. The Jones vector of the incident field and the Jones vector of the desired output fields must be known in order to create a metasurface for polarization conversion (Figure 2).  With this information, it is possible to determine a Jones matrix $J(x,y)$ for each spatial point $(x,y)$ on the metasurface plane connecting the incident and output waves.  Appropriate nanoantenna designs can be engineered to produce the calculated Jones matrix. Metasurfaces have been used to convert between linear and circular polarizations, between various linear polarization states, and between opposing circular polarization states [32,33,34]. Because circularly polarized input light is one of geometric-phase metasurfaces main limitations, it is necessary to precisely match the geometric phase to the propagation phase in order to obtain unrestricted control of polarization states. Typically, hybrid nanoantenna patterns are used to achieve this.

\subsection{OAM Control in Metasurfaces}

In order to create optical vortex beams with net topological charge  or OAM of light  ($l$), polarization-control in metasurfaces that permit manipulation of the propagation phase and the geometrical phase can also be used
 [30-47]. The helixed wave front of such OAM beams, which is a direct result of the dependence of phase on azimuthal angle, is one of its most distinctive features. The number of twists a wave-front has, determines the integer $l$ that represents the order of OAM (per unit wavelength). Since beams with different orders of OAM  are orthogonal they do not interfere with each other, widespread acceptance of OAM as an unrestricted degree of freedom of light that can be used in fast free-space optical communication systems has been achieved [39-48].

Spatial light modulators (SLM), holograms, laser mode conversion, and spiral phase plates (SPP) are examples of conventional techniques for producing OAM beams. On the other hand, by positioning nanoantennas with linearly increasing (or decreasing) phase shifts along the azimuthal direction, metasurfaces can produce helical wave fronts [49-55]. 
As a result, a  metasurface can  add an optical vortex to the incident light wave front, thus converting SAM into OAM,  this transformation is also termed spin-to-orbit conversion (SOC)~[32,33,34,70,71]. Due to the conservation of total angular momentum, the SOC typically facilitates the conversion of LHC and RHC polarization into states with opposing OAM (TAM). A metasurface can transform circular polarizations into states with independent values of OAM by adding an additional phase shift in the azimuthal direction. Recent years have seen the demonstration of the transformation of light with arbitrarily elliptical polarization states into orthogonal OAM vortex states [71,72].

\subsection{TAM Control in Metasurfaces}

It is possible to introduce a controlled geometric phase and establish a link between polarization (SAM) and phase using phase elements with spatially variable orientations. According to \cite{Beth36}, these devices are often constructed from periodic elements known as $q$-plates. The precise transformation made by $q$-plates can be written as \cite{DelvinScience, Mueller2017}:

\begin{eqnarray}
|L \rangle& \rightarrow & \exp[i2q \phi ]| R \rangle\\ \nonumber
|R \rangle& \rightarrow & \exp[-i2q\phi] | L \rangle,
\end{eqnarray}
where circular polarization LHC ($L$) and RHC ($R$) are mapped to its antiparallel spin state with an acquired OAM charge of $\pm 2q \hbar$ per photon. Spin-Orbit Conversion (SOC) is the common name for this transformation. As introduced in the previous Sections, the rotational elements (i.e., $\theta(x,y)$) are often the only ones that change the angle spatially. As a result, the OAM output states are limited to conjugate values ($\pm 2 q \hbar$). Additionally, the SOC operation carried out by $q$-plates is restricted to SAM states of circular polarization as a result of the device symmetry. A more general device is needed in order to perform the SOC operations of different SAM states, including elliptic polarization. As will be discussed later, $J$-plates can carry out this arbitrary mapping (Figure 2).

A $J$-plate (represented by the variable $J$) has the capacity to transfer two arbitrary TAM states into two arbitrary input SAM states, including but not limited to RHC or LHC. Any media that allows birefringence, absolute phase shift, and retarder orientation angles to vary spatially can be used to produce a $J$-plate. In other words, the relative phase shift between orthogonal spins ($ \phi= \phi(x,y)$) should vary spatially in addition to the fast axes of rotating plates. Metasurfaces were used to construct such $J$-plates  \cite{DelvinScience}. The $J$-plate's precise transformation is best described as (Figure 2):
\begin{eqnarray}
|\lambda+ \rangle& \rightarrow & \exp[im \phi ]|( \lambda +)^{*} \rangle\\ \nonumber
|\lambda - \rangle& \rightarrow & \exp[in\phi] | (\lambda -)^{*}\rangle,
\end{eqnarray}
where $|\lambda \pm \rangle$ are arbitrary elliptical polarization expressed as \cite{DelvinScience}:

\begin{equation}
|\lambda + \rangle=[\cos(\chi) , e^{i \delta} \sin(\chi)]^{T}, \hspace{1cm} |\lambda - \rangle=[-\sin(\chi) , e^{i \delta} \cos(\chi)]^{T},
\end{equation}
the parameters ($\chi, \delta$) determine the polarizations state. Implementing the SOC operation reduces to determining the actual Jones matrix $J(\phi)$ that transforms $J(\phi)|\lambda+ \rangle=e^{im\phi}|(\lambda +)^{*} \rangle$ and  $J(\phi)|\lambda - \rangle=e^{in\phi}|(\lambda - )^{*} \rangle$. It can be demonstrated that the required spatially varying Jones matrix has the form:

\begin{equation}
J(\phi)=e^{i \delta} 
\left[ \begin{array} {cc}
 e^{i \delta} ( e^{i m \phi} \cos(\chi)^2 +e^{i n \phi} \sin(\chi)^2) &    \frac{\sin(2\chi)}{2}(e^{im\phi}- e^{in\phi})\\ \nonumber
 \frac{\sin(2\chi)}{2}(e^{-im\phi}- e^{-in\phi})                               &  e^{-i \delta} ( e^{i m \phi} \sin(\chi)^2 +e^{i n \phi} \cos(\chi)^2)\
\end{array}
\right ].
\end{equation}

Unfortunately, the necessary control and sub-wavelength spatial variations in phase shift, birefringence, and orientation cannot be achieved with a conventional phase plate. On the other hand, sub-wavelength space metasurfaces and 2D metamaterials may provide such unprecedented control [56-69].

\section{SHEL in Photonic Materials: Experimental realizations}

\subsection{Experimental configurations}

To confirm the above theoretical predictions for SHE in tilted photonic materials, we performed a series of experimental measurements using the setups shown in Figure 3  \cite{GP1,GP2}. While in Ref. \cite{GP1} we use a sample of free-standing birefringent polymer foil, similar to the type Newport 05RP32-1064, as photonic material,  in Ref.  \cite{GP2}, we use a hyperbolic trench metamaterial structure. As a source of incident Gaussian beam, we employed a He-Ne laser (Melles Griot Griot 05-LHR-111) of wavelength $\lambda=633$ nm. The laser radiation was collimated using a microscope objective lens. We measure the anisotropic phase difference $\Phi_0$  versus the angle of the tilt $\vartheta$ via Stokes polarimetry methods \cite{GP1,GP2}. For this purpose we used the setup shown in Fig. 3 (a).\\

The appropriate linear-polarization state in the incident beam was chosen using the twin Glan-Laser polarizer (Thorlabs GL10) (P1). This was oriented at 45° polarization in the first experiment, equivalentely $\alpha=\beta=1/\sqrt2$ in the formalism introduced in the previous Sections. The Stokes parameters are then determined by passing the beam through the photonic material sample while utilizing a quarter wave plate (QWP) with a retardation angle of $\delta$ and a second polarizer P2 with a rotation angle of $ \gamma$, as shown in Fig. 3 (a). The Stokes parameters can be used to determine the phase difference using the expression:

 \begin{equation}
 \phi_0=\tan^{-1} (\frac{S_3}{S_2}),
 \end{equation}

\noindent where$S_2=I(0 ^{\circ}, 45^ {\circ})-I(0^{ \circ}, 135^{\circ})$ is the normalized Stokes parameter in the diagonal basis, and $S_3=I(90^{\circ}, 45^{\circ})-I(90^ {\circ}, 135^{ \circ})$ is the circular polarization's normalized Stokes parameter,  here the normalization factor $S_0$ is given by the total intensity of the beam. The measured phase using Eq. (5) is wrapped in the range $(-\pi, \pi)$. To calculate the unwrapped phase difference, we employ an unwrapping technique with a tolerance set to $0.001$ radians \cite{GP1,GP2}.

\begin{figure}[t]
\centering
\includegraphics[width=0.7\linewidth]{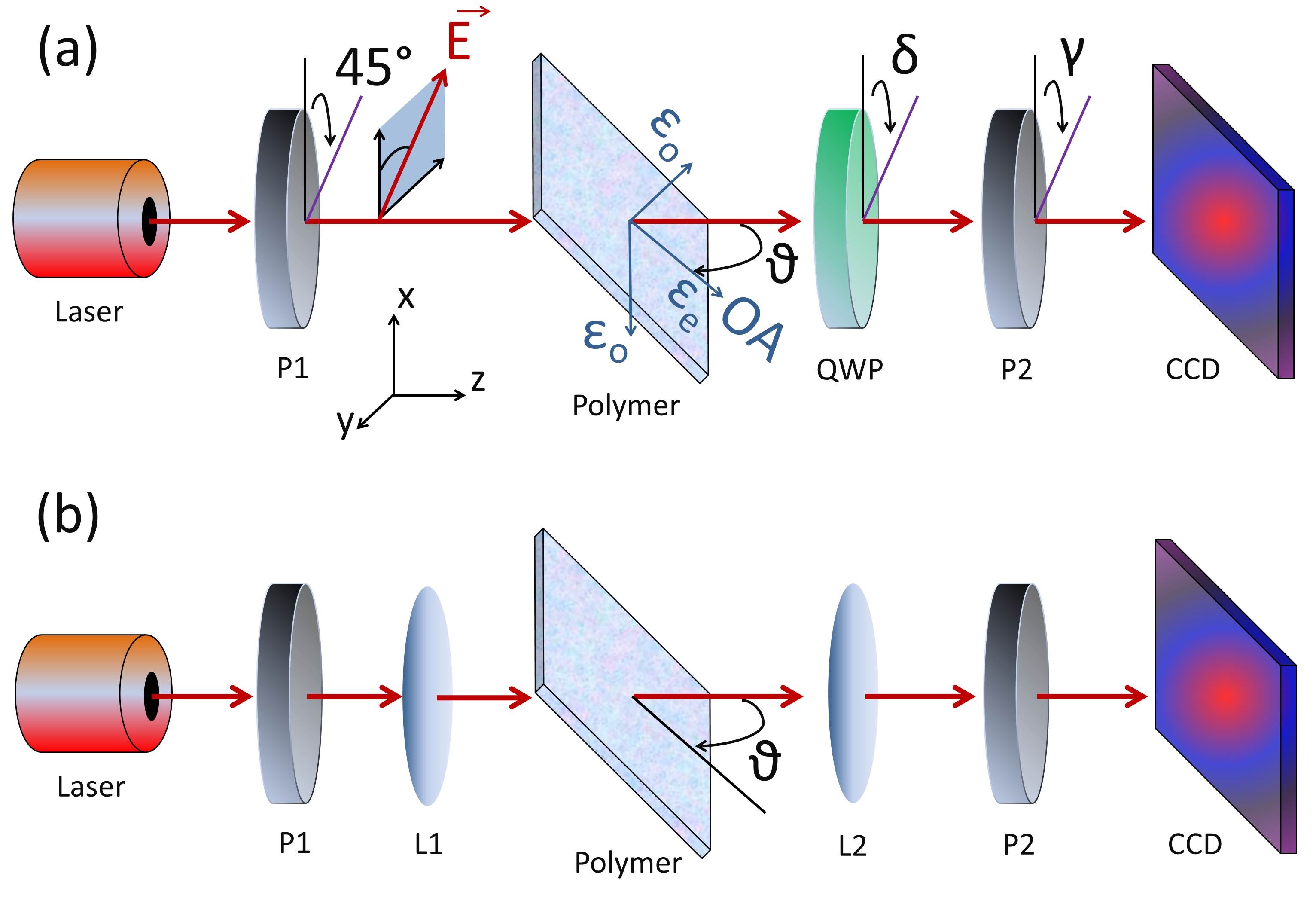}
\caption{Schematic depiction of experimental configurations for (a) polarimetric measurements and (b) quantum weak measurements. P1 and P2 represent double Glan-Laser polarizers (Thorlabs GL10), QWP is quarter wave plate. L1 and L2 are used to identify lenses. He-Ne (Melles Griot Griot 05-LHR-111) laser with a 633 nm emission wavelength; Thorlabs WFS150-5C CCD camera type. The ordinary and extraordinary permittivities of the photonic material sample, and their corresponding axes, are denoted as $\varepsilon_{o}$ and $\varepsilon_{e}$, respectively.}
\label{F1}
\end{figure}

\subsection{SHEL in Birrefringent Polymers}

In Ref. \cite{GP1}, a full experimental demonstration of SHE of light in birefringent polymers was provided. The schematic of tilted birefringent polymer film is depicted in Figure 4 (a).  We observe a spin-Hall effect via Stokes polarimetry ($k \langle \hat{Y} \rangle$) using a $50 \mu m$ polymer film which is 10 times larger than the shift observed in Ref. \cite{Bliokh2016}, for a 1000$\mu m$  Quartz sample. We attribute this increase to the polymers' greater effective birefringence. We also investigate the impact of tunable birefringence in the polymer film. By applying contrallable voltage, leading to tunable stress in the polymer, it is possible to induce  tunable birefringence in the polymer film, which can in turn induce controllable light shifts. In Figure 4 (b), we present numerical simulations of SHE for a set of stress-induced birefringence ranging from $ \Delta n=0.009, 0.03, 0.06, 0.07 $  as a function of tilting angle $ \vartheta$, maimal tunability is achieved at $\vartheta=0.3$ radians \cite{GP1}.

Additionally, using the quantum weak measurement apparatus shown in Fig. 3(b), we measured the spin Hall shift weakly and observed the quantum weak amplification effect. A CCD camera  (Thorlabs WFS150-5C) is used to image the beam. To achieve this, two lenses (L1) and (L2) with a focal distance of $f=6$ cm were implanted. Pre-selected and post-selected polarization states are produced by polarizers Glan-Thompson Polarizers P1 and P2, respectively, having polarization states of $|\psi \rangle$ and $|\psi' \rangle$. The first lens (L1), which had a 6 cm focal length, generated a Gaussian beam with a waist of 30 $\mu$m and a Rayleigh range of $z_{R}=4.6$  mm. Therefore, $z/z_{R}=10.86$ is the propagation amplification factor for a CCD camera placed at a distance of $z=5$ cm. The amplification factor due to crossed polarizers results $1/\epsilon \approx 1.83 \times 10^{-2}$. For $k=\frac{2 \pi}{\lambda}$, the overall weak amplification factor becomes $A = \frac{z}{z_{R}} \times  \frac{1}{k \epsilon}$ = 200, this is confirmed in the experiment which involved a displacement between centroids of $\Delta Y = 1000 \mu$m  between post-selection polarizers oriented  at $\epsilon=-1/1.83 \times 10^{-2}$ (Fig. 4(c) Top), and $\epsilon=+1/1.83 \times 10^{-2}$ (Fig. 4(c) Bottom) is measured at a tilt angle $\vartheta=20$ $^{\circ}$, consequently, the SHE is amlified by a factor $A = 200$. For crossed polarizers ($\epsilon=0$), a Hermite-Gaussian distribution is created from the input Gaussian beam (Fig.4(c) Middle), and the two centroids are roughly separated from one another by approximately $\Delta Y=1000\mu m$.  

In conclusion, we experimentally demonstrated the fine lateral circular birefringence of a tilted birefringent polymer, the first instance of the SHEL in a polymer material \cite{GP1}. We revealed experimental results of this nanometer-scale phenomena and found a quantum weak amplification factor of 200 using Stokes polarimetry and quantum-weak measurement techniques. Because the polymer's birefringence may be controlled using mechanical stress in the case of stress-induced birefringence or voltage in the case of liquid crystals, this lateral shift could be utilized as an optical switch at the nanoscale scale. Numerous cutting-edge applications in photonics, nano-optics, quantum optics, and metamaterials might become possible as a result. Such emergent application of SHEL are discussed in the following Sections. 

\begin{figure}[h!]
\centering
\includegraphics[width=0.8\linewidth]{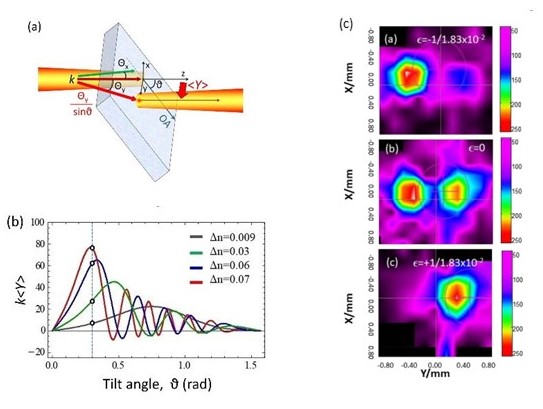}
\caption{(a) Paraxial beam's transmission 
through a clear birefringent polymer slanted layer as shown in 3D geometry. The beam goes through
transversal shift $<\hat{Y}>$  at the nanometer scale brought on by the Spin Hall Effect (SHE) of light.
The paraxial angles ($\Theta_{x}, \Theta_{y} $) identify the propagation direction of
the incident beam's wave vectors $k$.  
(b) A tunable birefringent polymer results in enhanced SHE of light as shown in numerical simulations considering different birefringence:
$\Delta n$  0.009 (grey, index difference in quartz), $\Delta n$   0.03 (green),
$\Delta n$   0.06 (blue), and $\Delta n$  0.07 (red, index difference in stretched
polymers [32]). Vertical dashed line at $\nu$   0.3 (rad) is to
show the tunability of beam shift by different birefringence. (c) Transverse intensity distributions (a.u.) in false scale for an o-polarized beam transmitted through a tilted polymer plate and post-selected in the almost e-polarized state,  with a tilt angle $\vartheta=20^{\circ}$. Top: Post-selected polarization state with $\epsilon=-1/83 \times 10^{-2}$. 
The beam centroid is shifted, resulting in a measurement of weak value $<\hat{Y}>_{weak}=-500 \mu m$. Middle: With crossed polarizers ($\epsilon=0$), a Hermite-Gaussian intensity distribution with peaks spaced apart from one another by $\Delta Y=1000\mu m$ is created from a Gaussian distribution, corresponding to a weak  amplification factor $A=200$. Bottom: Post-selected polarization state with $\epsilon=1/83 \times 10^{-2}$, which corresponds to a weak value measurement of $<\hat{Y}>_{weak}=+500 \mu m$.}
\label{F1}
\end{figure}

\subsection{SHEL in Metamaterials}

In Ref. \cite{GP2}, we present experimental findings for SHE of light in Hyperbolic Metamaterials (HMM).  In particular, we experimentally show enhanced spin-Hall effect of light in HMMs in the visible region and exceptional angle sensitivity. As shown in Fig. 5 (a), the effect is shown in the transmission configuration using an HMM that is a few hundred nanometers thick and made up of alternating layers of metal and dielectric as illustrated in Fig.  5 (a). With a change of $\approx$ 0.003 rad ($\approx$ 0.17 deg) in the angle of incidence, the transverse beam shift in the HMM configuration can alter dramatically, from essentially no beam shift to a few hundred microns. Therefore, it is to be expected that compact spin Hall photonic devices can take advantage of the huge photonic spin Hall enhancement in such a tiny structure, with a great angular sensitivity to manipulate photons via polarization.\\

The HMM sample has eight gold-alumina periods placed on a 500 $\mu$m thick glass substrate for a total thickness of 176 nm.  Al$ 2$O$ 3$(10 nm)—APTMS(1 nm)—Au(10 nm)—APTMS(1 nm)  are the four layers that make up one period of the HMM structure.  Amino Propyl Tri Methoxy Silane, often known as APTMS, is an almost loss-free adhesion layer that is beneficial for highly confined propagating plasmon modes \cite{GP2}. Al$ 2$O$ 3$ layer was deposited via atomic layer deposition, upon Au layer sputtering. Calculated as \cite{30} are the HMM's ordinary and extraordinary permittivities, indicated as $\varepsilon_{o}$  and $\varepsilon_{e}$, respectively. 

Based on the effective media approximation \cite{GP2}, HMMs are treated as homogenized uniaxial mediums with effective permittivities.
The thicknesses of individual layers are assumed to be deeply subwavelength within the effective media approximation \cite{GP2}. The wavelength range of $\lambda$  = 500–700 nm, normalized by the unit cell $\Lambda$  = 22 nm, yields the ratio of $\Lambda/\lambda$ = 1/22.7 — 1/31.8  Therefore, it is justified to use the effective media approximation in our scenario.

Calculated as described in \cite{30}, the HMM's ordinary and extraordinary permittivities, are indicated as $\varepsilon {o}$  and $\varepsilon {e}$, respectively:

\begin{equation}
\varepsilon_{o}=f_{Au}\cdot\varepsilon_{Au}+f_{Al_2O_3}\cdot\varepsilon_{Al_2O_3}+f_{APTMS}\cdot\varepsilon_{d_{APTMS}}
\end{equation}

\begin{equation}
\varepsilon_{e}=(\frac{f_{Au}}{\varepsilon_{Au}}+\frac{f_{Al_2O_3}}{\varepsilon_{Al_2O_3}}+\frac{f_{APTMS}}{\varepsilon_{APTMS}})^{-1},
\end{equation}

\noindent where $f_{m}$ and $f_{d}$ are the volume fractions of metal and dielectric, respectively, and $\varepsilon_{m}$ and $\varepsilon_{d}$ are the permittivities of metal and dielectric, respectively. The Drude model with the thickness-dependent correction is used to describe the Au film's permittivity, $\varepsilon m$ \cite{31}. 

APTMS has a refractive index of 1.46 \cite{32}.   The dispersion of the HMM effective permittivities in the visible region is seen in Fig. 5 (b).   With $\lambda$= 500 nm as the zero crossing wavelength for our HMM structure, type II HMMs ($\varepsilon_o <$ 0 and $\varepsilon_e >$ 0) are formed when red wavelengths are approached.


We ran simulations based on the theory  developed by T. Tang \textit{et al. } \cite{GP2}  in order to comprehend how the spin Hall beam shift behaves in HMMs,  with realistic parameters (Fig. 5 (b)). Air serves as the ambient medium for the entire HMM-SiO$ 2$ substrate structure, as depicted in Fig. 5 (a) . We consider that the incident light is impinging on the HMM structure with an incident angle $\theta_{i}$ in the y-z plane. Since $\epsilon 1=\epsilon 2=\epsilon 5$ = 1 (air), $\epsilon 4$  corresponds to the SiO$ 2$ substrate, and  $\epsilon 3$ corresponds to the HMM, the relative permittivities of the media in regions 1–5 are indicated by $\epsilon_{i}$ $(i = 1, 2, 3, 4, 5)$ respectively. The HMM is considered to be uniaxially anisotropic, non-magnetic, and to have a relative permittivity tensor ($\epsilon_3$):

\begin{equation}
\epsilon_3 = 
\left(\begin{array}{ccc} \epsilon_{o} & 0& 0\\ 0 & \epsilon_{o} & 0  \\
0 & 0 & \epsilon_{e}
\end{array}\right)
\end{equation}

Taking into account the input waist Gaussian beam $\omega_0$:

\begin{equation}
E_{H,V}(x,y)=\frac{\omega_0}{\sqrt {2 \pi}}\exp{-\frac{\omega_0^2(k_{x}^2+k_{y}^2)}{4}},
\end{equation}

we can define the transverse beam shifts after transmission through the structure in the form:
\begin{equation}
\eta_{H,V}^{\pm}=\frac{x|E_{H,V}|^{\pm}}{|E_{H,V}|^{\pm}},
\end{equation}

\noindent where the transverse shifts for the right hand circular (RHC) and left hand circular (LHC) polarizations are indicated by $\eta^{\pm}$, respectively. The terms for the transverse shifts contain z-dependent and z-independent terms, which, respectively, represent angular and spatial transverse shifts \cite{33}. In this case, our attention is on the spatial transverse shift of light transmitted by the HMM waveguide, which has the following form:

\begin{equation}
<\hat{Y}>=\pm \frac{k_1 \omega_0^2(t_{s}^2\frac{\cos (\theta_{t})}{\sin (\theta_{i})}-t_{s}t_{p}\cot (\theta_{i}))}
{k_{1}^{2}\omega_0^2t_{s}^2+\cos^2(\theta_{i})(t_{s} \frac{\cos (\theta_{t})}{\cos(\theta_{i})}-t_{p})^2+ (\frac{d t_{s}}{d \theta_{i}} )^2  },
\end{equation}

where $t_{s,p}$ are the transmission amplitudes for the $s,p$ modes, respectively \cite{GP2},  $\theta_{t}$ is the transmission angle, and $k {1}=n 1 k=n 1\frac{2 \pi}{\lambda}$ with $n 1=1$ (air) are all constants. We assume transmission along the laser beam axis, thus $(\frac{d t {s}}{d \theta {i}}) \approx 0$ for a large beam waist and $\theta t$ = 0f or a large beam waist.


We carried out a number of characterizations with the help of the polarimetric setup depicted in Fig . 3 (a) to highlight the angular sensitivity of the photonic SHE in the HMM structure. Fig. 5 (a). Polarimetric and quantum weak measurement techniques can be used to determine the transverse beam shift \cite{GP1,GP2, 22}. We used a He-Ne laser with a wavelength of $\lambda = 633$ nm and a diode laser with a wavelength of $\lambda =520$ nm, as sources of an incident Gaussian beam. A microscope objective lens was used to collimate the laser light. Using Stokes polarimetry, we calculated the anisotropic phase difference $\Phi 0$  vs angle $\theta i$. A Quarter Wave Plate (QWP) and then a Glan-Thompson polarizer (P1) are used to produce the input polarization state (RHC). The Stokes parameters can be used to determine the phase difference using the expression:

\begin{equation}
\Phi_0=\arctan (S_3/S_2),
\label{phase}
\end{equation}

\noindent where the normalized Stokes parameter for circular polarization is defined as $S_{3}=I(90 ^\circ,45 ^\circ)-I(90 ^\circ,135 ^\circ)$, and the normalized Stokes parameter for diagonal basis is defined as  $S_{2}=I(0 ^\circ,45 ^\circ)-I(0 ^\circ,135 ^\circ)$, where normalization factor $S_0$ is determined by the total beam intensity. The retardation angle of the quarter wave plate QWP2 and the rotation angle of the polarizer P2 are represented by $\delta$ and $\alpha$ of $I(\delta,\alpha)$, respectively. The range (-$\pi,\pi$). encompasses the measured phase obtained using Eq. (5). We employ an unwrapping algorithm \cite{GP1} to calculate the unwrapped phase difference  with a 0.01 rad tolerance set.

The transverse beam shift, $<\hat{Y}>$, is found from the measured phase, $\Phi_0$,  
\begin{equation}
<\hat{Y}> = \frac{1}{k}\cot(\pi/2-\theta_i)[-\sigma(1-\cos\Phi_0)+\chi \sin\Phi_0],
\label{eq_kY}
\end{equation}
\noindent where the Stokes parameters for the circular and diagonal bases, respectively, are $\sigma$ and $\chi$. They are given by $\sigma$ = 2Im($\alpha^*\beta$), and $\chi$ = 2Re($\alpha^*\beta$) from the Jones vector $|\psi\rangle=(\alpha, \beta)$ of the incident beam, respectively. The incident beam has $\alpha= \frac{1}{\sqrt{2}},\beta=\frac{i}{\sqrt{2}}$, which corresponds to RHC polarization.


 The experimental beam shift points for wavelengths $\lambda$ = 520 nm and $\lambda$ = 633 nm are contrasted with the findings of simulations that account for the HMM's losses in Figure 5(c) and (d). $\varepsilon_o$ = -2.9460 + 1.2163i and $\varepsilon e$  = 5.5063 + 0.3399i for $\lambda$ = 633 nm and $\varepsilon o$= -0.3257 + 1.3664i and $\varepsilon e$= 5.8847 + 1.9963i for $\lambda$ = 520 nm, respectively, are the effective permittivities.
  
In Figures 5, we can see the characteristics of the photonic spin Hall effect in transmission-configured HMM systems. The tranverse beam shift in HMMs is quite sensitive to the incident angle; for $\lambda$ = 633 nm, changes from $\theta i$  = 0 rad to just $\theta i$  = 0.003 rad (0.17 $circ$) [Fig. 5(c)] causes a massive beam shift of several hundred microns while demonstrating milliradian-level sensitivity. The angular variation from $\theta i$ = 0.003 rad to above also significantly changes the beam shift from $<\hat{Y}>$ = 105 $\mu$m to merely $<\hat{Y}>$ = 10 $\mu$m, which is almost one order of magnitude difference. Given that we observed the same result in dielectric media \cite{22}, the significant anisotropy of HMMs is believed to be the cause of the beam shift's strong peak. With a wider beam diameter and  wavelength $\lambda$ = 520 nm (Fig. 5(d)). The peak shift of $<\hat{Y}>$ = 270 $\mu$m is reached by the incident angle change of 0.001 rad (0.057 $^\circ$) only ($\approx$ 4700 $\mu$m/$^\circ$). This means that the beam shift exhibits even sharper resonance and, as a result, greater angular sensitivity. The beam shift dramatically decreases to $<\hat{Y}>$  = 10 $mu$m and less when the incident angle increases, for example, $\theta_i$ = 0.01 rad (0.57 $^\circ$).

The simulation results, which account for alignment error and beam divergence, are quantitatively consistent with the experimental results. According to estimates, the He-Ne laser's beam divergence is approximately $\Delta\theta i$ = 0.01 rad (0.57 $^\circ$) while the green diode laser's beam divergence is approximately $\Delta\theta i$ = 0.02 rad (1.14 $^\circ$). Additionally, the rotating mount's inaccuracy of around 0.0017 rad (0.1 $^\circ$) limits the angular resolution. Given that $\lambda$ = 633 nm,  the experimental transverse beam shift under normal incidence approaches $<\hat{Y}>$ = 50 $\mu$m (Fig, 5 (d)).

However, the transverse beam shift drastically decreases to $<\hat{Y}>$ = 5 $\mu$m when the incidence angle is inclined by just $\theta i$ = 0.035 rad ($\approx$ 2 $^\circ$).  Such findings show the great angular sensitivity of the SHE in HMMs and prove that the transverse beam shift may be controlled within an order of magnitude range by minor angular adjustments. Experimental measurements were utilized to determine the waist ($w_0$ = 100 $\mu$m) for the simulations.

According to Figure 5 (d), with $\lambda$  = 520 nm, we observe $<\hat{Y}>$= 165 $\mu$m for $\theta i$ = 0 rad . At $\theta i$ = 0.035 rad ($\approx$ 2 $^\circ$), the beam shift decreases to $<\hat{Y}>$= 5 $\mu$m. The previously observed spin Hall effect in dielectric anisotropic media, such as a quartz crystal ($\approx$ 150 $\mu$m / 5 $^\circ$ = 30 $\mu m/^\circ$)  and polymer film (approximately 250 $\mu$m / 20 $^\circ$ = 12.5 $\mu m/^\circ$) \cite{22}, is strikingly different from the extremely high angular sensitivity of this new device. When the beam shift is two orders of magnitude larger and occurs within a few degrees of sample tilting. Additionally, the HMM has a thickness of only 176 nm as contrasted to the dielectric materials, which have thicknesses of $50 \mu m$ for polymer films and $1 mm$  for quartz plates.

In summary, we experimentally proved the photonic SHE in a hyperbolic metamaterial at visible wavelengths for the first time. The incidence angle has a significant impact on the tranverse beam shift in the transmission arrangement. We found that a little difference of a few milliradians can affect the beam shift by two orders of magnitude, going from a few hundreds of microns to a few microns, for example. A HMM that is two hundreds of nanometers thick achieves this tremendous angular tunability. Such sensitivity can result in small and compact spin Hall devices, such as switches, filters, and sensors, that control light at the nanoscale by varying the wavelength, incidence angle, and spin.

\begin{figure}[h!]
\centering
\includegraphics[width=0.8\linewidth]{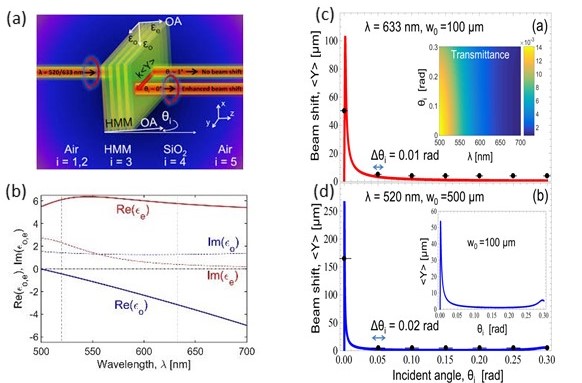}
\caption{(a) Scheme of the Spin Hall Effect (SHE) of light in hyperbolic metamaterials. The incidence angle, indicated by $\theta_i$, switches the transverse beam shift along the y-axis denoted as $<\hat{Y}>$. The HMM structure's unit cell is made up of Al$_2$O$_3$(10 nm)-APTMS(1 nm)-Au(10 nm)-APTMS(1 nm). Eight periods totaling 176 nm in thickness make up the HMM. (b)According to the effective media approximation, the effective permittivities of a multilayer HMM structure made up of a unit cell of Al$ 2$O$ 3$(10 nm)-APTMS(1 nm)-Au(10 nm)-APTMS(1 nm) were calculated. The structure exhibits type II ($\varepsilon o <$ 0 and $\varepsilon e >$ 0)  HMM behavior in the visible wavelength range. The vertical dashed lines indicate the wavelengths $\lambda$ = 520 nm and 633 nm at which experiments were conducted.  Measured and calculated spin Hall transverse shifts $<\hat{Y}>$  for (c) $\lambda$  = 633 nm and (d) $\lambda$  = 520 nm, respectively, are shown in solid lines. Beam waist $w o$ is fitted to the simulation results in (a) and (b). Note that there is an estimated beam divergence of $\Delta\theta i$ = 0.01 rad for $\lambda$ = 633 nm and $\Delta\theta i$ = 0.02 rad for $\lambda$ = 520 nm indicated as lateral error bars, respectively. Further detalis are in the text}
\label{F1}
\end{figure}

\section{Emergent Applications of SOI in Photonic Materials}

SOIs may be engineered to modify they way in which an artificial substance disperses. SAM and OAM can therefore both be used as a means of control for light. The multifunctional spin-dependent element is made available by the geometric phase design, which also permits the spin-based optical devices. In addition, unlike conventional devices based on dynamic phase, the manipulation of cumulative geometric phase is essentially the control of light polarization. In conventional settings, a phase distribution is created by varying the reflective indexes or the thickness of optical materials. However, planarization and miniaturization of components are necessary for the advancement of integration optics. The photonic SHE may be developed in conjunction with the functions of conventional components to create infinitesimally tiny and multifunctional devices. Photonic SHE devices that provide the basic optical components variation for emergent applications of SOIs will be briefly outlined in this section.

\subsection{Sensing mechanical and optical properties of 2D Materials and Metasurfaces}

The transverse and spin-dependent shifting of light is referred to as the optical Spin Hall Effect (SHE). As a result, it has been enthusiastically advocated for a variety of sensing applications, including biosensing, material interface studies, polarization-dependent sensors, and refractive index spectroscopy  [83, 84, 85]. SHE biosensors usually include a graphene sheet, or some alternative 2D material, an Au film, and a BK7 glass. A change in the concentration of biomolecules in the sensing medium will cause a localized change in the refractive index close to the graphene surface. The photonic SHE's spin-dependent shifs are sensitive to changes in the sensing medium's refractive index.
More specifically, it is possible to infer a quantitative link between the spin-dependent splitting and the sensing medium's refractive index. Furthermore, by varying the refractive index of the sensing medium, the spin-dependent splitting in the biosensor may be studied. Similar SHE-based sensing applications have been documented \cite{scirep2018}  (Fig. 6(a) and (c)). Weak measurement amplification methods can be used to further improve such sensing applications.

\subsection{Spintronics based on photonic  SHE}

The creation of a spin current perpendicular to the direction of the charge current flow constitutes the Spin Hall Effect itself. The optical SHE based on polaritons,  and consisting of a separation between real space and momentum space, was recently predicted for laser-induced spin-polarized exciton-polaritons, in a semiconductor microcavity, due to a combination of structural disorder-based dispersion of exciton-polaritons, and an effective magnetic field resulting from polarization splitting of the polariton states. The excitonic spin current is controlled by the linear polarization of the laser pump. The first experimental evidence for this effect was reported in [86, 87, 88]. 
It was stated that polariton spin currents might travel over 100 $\mu$m distances in a superior GaAs/AlGaAs quantum microcavity. By rotating the laser pump polarization plane, it is feasible to switch the spin currents directions, opening the door to a host of amazing applications in optical spin switching and spintronics, in addition to further emergent spin-based metrology functionalities (Fig. 6 (d)).

\subsection{Applications in Quantum Information Networks}

A polarizing beam splitter or spin-dependent splitter can separate the orthogonally polarized components of a beam into different propagation directions. Since the beam is divided into several spatial modes according to its polarization, the photonic SHE generator may intuitively be thought of as a polarization beam splitter. Metasurfaces have recently been shown to be able to replace bulk optical components and be extended into the single-photon quantum optical regime. For use in quantum network applications, the decreased propagation losses brought on by metasurfaces make it possible to realize a spin-dependent splitter at the single photon level. In particular, SHE switches and decoders are a fundamental building block for Quantum Information Networks where polarization dependent operations, such as C-NOT and SWAP gates, are daily required  [79, 81, 82] (Fig. 6 (b)).

\begin{figure}[h!]
\centering
\includegraphics[width=0.8\linewidth]{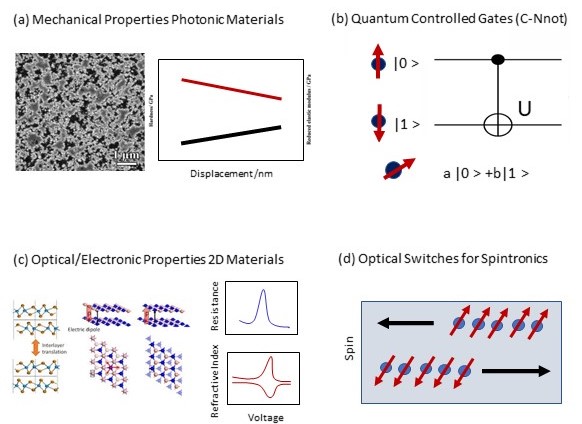}
\caption{SHE of light applications in precision measurements of (a) mechanical properties of photonic materials, (b) quantum network applications, (c) optical/electronic properties of 2D materials, (d) optical switching for spin-based metrology and spintronics. Further details are in the text. }
\label{F1}
\end{figure}

\subsection{Photonic Multiplexing and Muticasting Devices}

Metasurfaces can contribute to a new class of on-chip scaleble devices, which are expected to revolutionize nanophotonic and optoelectronic circuitry through smart integration of multiple functions in metallic, dielectric, or semiconductor building blocks. Metasurfaces among artificially produced materials offer considerable promise to succeed in technologically relevant applications because of their 2D nature, which is a key benefit for wafer-scale manufacturing and integration. Here, we will concentrate on applications of metasurfaces for high-speed data transmission in integrated OAM multiplexing and multicasting devices [79-81].

In order to distinguish between the many orthogonal channels, OAM division multiplexing is an experimental technique for boosting the transmission capacity of electromagnetic signals, as stated in [82]. It corresponds to transmission over a few kilometers in OAM-maintaining fibers equivalent to wavelength division multiplexing (WDM), temporal division multiplexing (TDM), or polarization division multiplexing (PDM). The extremely bulky optical components needed for OAM generation and OAM detection are one of the key constraints for scalable OAM multiplexage in addition to the lower transmission range [33], in the region of space that OAM of light covers. While OAM multiplexing can access a theoretically endless collection of states and as a result can offer an infinite number of channels for multiplexage, SAM or polarization multiplexing only offers two orthogonal states that correspond to the two states of circular polarization. 

Although 2.5 Tbit/s transmission rates in MIMO systems have been reported, OAM multiplexing still remains an experimental approach and has only been tested in the lab so far, over relatively short distances of a few Km over OAM maintaining fibers. Nevertheless, it promises very significant improvements in bandwidth. The extremely bulky optical components needed for OAM generation and OAM detection are one of the key constraints for scalable OAM multiplexage in addition to the lower transmission range. Recently [73-77], the first experimental demonstration of an OAM multiplexing technique based on single-layer metasurfaces in the Terahertz (THz) band was realized. In partiuclar, OAM multiplexing with four channels is made possible by the developed structure's ability to produce four focused phase vortex beams with various topological charges (or OAM number $l$) when illuminated by a Gaussian beam.

The OAM signal is demultiplexed when a single vortex beam is employed as the incident light because only one channel is recognized and extracted as a focal spot. The subwavelength-level thickness of the metasurface structure expands the range of viable methods for the integration and downsizing of THz communication systems. Excellent agreement between theoretical predictions and practical results can be seen in the performance of the developed OAM multiplexing and demultiplexing device, proving its suitability for scaled ultra high-speed THz communications.

\section{Conclusions}

In conclusion, we presented a thorough review of recent developments in Spin Orbit Interactions (SOIs) of light in photonic materials. In particular, we highlighted progress on detection of Spin Hall Effect (SHE) of light in hyperbolic metamaterials and metasurfaces via polarimetric measurements, reporting unprecendented angular resolution at visible wavelength. Moreover, we outlined some fascinating future directions for emergent applications of SOIs of light in photonic devices of the upcoming generation. As a rapidly expanding interdisciplinary field, SOIs of light in 2D metamaterials and metasurfaces has important emergent applications in nanophotonics, biosensing, plasmonics, quantum optics, and telecommunication. SOIs in metamaterials and metasurfaces largely guarantees exceptional performance and versatility in the exact control of optical fields. Moreover, applications made possible by optical metasurfaces  decreased dimensionality are significantly different from those made possible by bulk metamaterials. In general, metasurfaces can offer a novel tool for scalable OAM generation and conversion with minimal losses. This feature can encourage many applications in integrated on-chip OAM generation, such as multiplexing and multicasting approaches, which may hold the promise of boosting transmission capacity and resolving scalability problems beyond the state of the art  [12, 13, 77, 78, 79, 80, 81, 82].

\section{Acknowledgements}
This Review is intended as a contribution to the advancement of scientific knowledge, for the benefit of the entire society, and its future generations. The author is  grateful to Konstantin Bliokh, Andrei Lavrinenko, and Ricardo Depine for many helpful discussions. The author ackowledges Osamu Takayama and Radu Malureanu for providing the metamaterial samples used in Ref. \cite{GP2}. This work was supported by ANPCyT via grant PICT Startup 2015 0710 and UBACYT PDE 2017.

\end{document}